\documentclass[12pt]{article}
\usepackage{amssymb}

\usepackage{graphicx}

\ifx\pdfoutput\undefined
    \relax
\else
    \usepackage{epstopdf}
\fi

\makeatletter
\def\numberbysection{\@addtoreset{equation}{section}
 	\def\theequation{\thesection.\arabic{equation}}}
\makeatother

\numberbysection

\newcommand{\be}{\begin{eqnarray}}
\newcommand{\ee}{\end{eqnarray}}
\newcommand{\non}{\nonumber}

\newcommand{\ch}{\mathop{\rm cosh}\nolimits}
\newcommand{\sh}{\mathop{\rm sinh}\nolimits}

\newcommand{\csch}{\mathop{\rm cosech}\nolimits}

\newcommand{\h}{\ensuremath{\mathsf{h}}}
\newcommand{\sgn}{\mathop{\rm sgn}\nolimits}

\newcommand{\hh}{\mathop{\mathcal H}\nolimits}

\begin{document}

\begin{titlepage}
\strut\hfill UMTG--253
\vspace{.5in}
\begin{center}

\LARGE Finite-size correction and bulk hole-excitations for special case
of an open XXZ chain with nondiagonal boundary terms at roots of unity\\[1.0in]
\large Rajan Murgan\\[0.8in]
\large Physics Department, P.O. Box 248046, University of Miami\\[0.2in]  
\large Coral Gables, FL 33124 USA\\

\end{center}

\vspace{.5in}

\begin{abstract}
Using our solution for the open spin-$1/2$ XXZ quantum spin chain with $N$ spins and 
two arbitrary boundary parameters at roots of unity, the central charge and 
the conformal dimensions for  bulk hole excitations are derived from the $1/N$ 
correction to the energy (Casimir energy).
     
\end{abstract}
\end{titlepage}

\setcounter{footnote}{0}

\section{Introduction}\label{sec:intro}

The integrable open spin-$1/2$ XXZ chain has been subjected to intensive studies
due to its growing applications in various fields of physics, e.g., statistical 
mechanics, string theory and condensed matter physics. However, obtaining 
exact solutions for this model has been a rather 
challenging and elusive task for many years. Various progress have been 
made in obtaining solutions for this model, either using the Bethe ansatz approach for diagonal \cite{Gaudin}--\cite{Sklyanin},
constrained nondiagonal \cite{Ne1}--\cite{YNZ} and nondiagonal cases at roots of unity \cite{MN1}--\cite{MNS3}, 
or using the representation theory of the $q$--Onsager algebra for general nondiagonal cases \cite{Bas}. 
Approaches based on boundary Temperley-Lieb algebra and its representations have also
been presented recently, from which the spectral properties
of the chain have been studied \cite{DNPR}.
Upon obtaining the desired solution, the next natural question that needed to be 
addresed is its practicality within various contexts. One 
important area where these solutions have found creditable applications is in 
determining finite size corrections to the ground state energy. By relating
to conformal invariance, these finite size corrections are shown to be 
related directly to other crucial parameters like the critical indices, 
central charge and conformal dimensions \cite{Affleck}--\cite{Cardy}. There are few methods and approaches 
to accomplish this task. De Vega and Woynarowich \cite{dVW} derived integral equations 
for calculating leading finite-size corrections for models solvable by Bethe Ansatz approach \cite{Hamer2}.
This was then generalized to nested Bethe Ansatz models as well \cite{dV}. Another approach was 
introduced by Woynarowich and Eckel \cite{WE,W}, which utilizes Euler-Maclaurin formula 
and Wiener-Hopf integration to compute these corrections for the closed XXZ 
chain. Others have also studied more general integrable spin 
chain models e.g., XXZ diagonal \cite{Alcaraz,Hamer}, nondiagonal cases \cite{AN}, 
quantum spin $1/2$ chains with non-nearest-neighbour short-range interaction 
\cite{ID} and XXZ($1/2, 1$) which contains alternating spins of $1/2$ and $1$ \cite{DM}, 
within similar framework. Other approaches e.g., based on NLIE (Nonlinear Integral Equations) have also
been successful in determining these effects for integrable lattice models \cite{Klumper} and
related integrable quantum field theories, such as the sine-Gordon model with periodic \cite{DesDevega}--\cite{Feverati2},
Dirichlet \cite{LeClair}--\cite{ANS} and Neumann boundary conditions 
\cite{AN,AhnZoltan}. 
   
With similar aim in mind, utilizing an exact solution for the integrable spin-$1/2$ XXZ chain
with nondiagonal boundary terms at roots of unity we found earlier for even number of sites \cite{MN2,MNS2}
\footnote{This solution, in contrast to \cite{Ne2}--\cite{YNZ} does not assume any constraint among the boundary parameters.},
and extending the solution to account for odd number of sites as well, we 
compute the correction of order $1/N$ (Casimir energy) to the ground state energy together with its low lying excited states (multi-hole states).
We employ the method introduced by Woynarovich and Eckle \cite{WE} that makes use of
Euler-Maclaurin formula \cite{WW} and Wiener-Hopf integration \cite{MF}. In particular, we compute
the analytical expressions for central charge and the conformal dimensions of low lying excited states.  
We also compare these analytical results to corresponding numerical results obtained 
by solving the model numerically for some large number of sites.

The outline of this article is as follows. In Section 2, we review the Bethe Ansatz solution \cite{MN2,MNS2}. 
We also present an extension of that result to include solution for
odd $N$. In Section 3, we present the calculation of $1/N$ correction to the 
ground state energy and hence our results for the central charge and 
conformal dimensions of low lying excited states. 
We notice that the lowest energy state for even $N$  of this model has one 
hole. Hence, the true ground state (lowest energy state without holes) 
lies in the odd $N$ sector. Similar behaviour are also found for the open chain
with diagonal boundary terms, for certain values of boundary parameters \cite{Nepprivate}. 
It is known that (critical) XXZ model with nondiagonal boundary terms 
corresponds to (conformally invariant) free Boson with Neumann boundary condition
whereas the diagonal ones are related to the Dirichlet case \cite{AhnBellacosa,Saleur2,Affleck2,AhnZoltan}. 
Although the model we study here has nondiagonal boundary terms, we find that 
the conformal dimensions for this model resemble that of the Dirichlet boundary condition. 
Some numerical results are presented in Section 4 to confirm and support the analytical results derived in Section 3.
Here, we solve the model numerically for some large but finite $N$ and further
employ an algorithm due to Vanden Broeck and Schwartz \cite{VBS}--\cite{Hamer3} 
to extrapolate the results for $N\rightarrow\infty$ limit. We conclude with a 
discussion of our results and some potential open problems in Section 5.
   
\section{Bethe Ansatz}\label{sec:Bethe}

We begin this section by reviewing recently proposed Bethe Ansatz solution \cite{MN2,MNS2} for 
the following model \cite{dVGR,GZ}
\be
\hh &=& \hh_{0}
+ {1\over 2}\sh \eta \big( 
 \csch \alpha_{-}\sigma_{1}^{x} 
 + \csch \alpha_{+}\sigma_{N}^{x}
\big)  \,, \label{Hamiltonian} 
\ee
where the ``bulk'' Hamiltonian is given by
\be 
\hh_{0} = {1\over 2}\sum_{n=1}^{N-1}\left( 
\sigma_{n}^{x}\sigma_{n+1}^{x}+\sigma_{n}^{y}\sigma_{n+1}^{y}
+\ch \eta\ \sigma_{n}^{z}\sigma_{n+1}^{z}\right) \,. 
\label{bulkHamiltonian}
\ee
In the above expressions, $\sigma^{x}$, $\sigma^{y}$, $\sigma^{z}$ are the usual
Pauli matrices, $\eta$ is the bulk anisotropy parameter (taking values $\eta = {i\pi\over{p+1}}$, with $p$ odd),
$\alpha_{\pm}$ are the boundary parameters, and $N$ is the number of spins/sites. 
Note that, this model has only two boundary parameters. Other boundary parameters
(as they appear in the original Hamiltonian in \cite{dVGR}) have been set to 
zero. We restrict the values of $\alpha_{\pm}$ to be pure imaginary 
to ensure the Hermiticity of the  Hamiltonian. The Bethe Ansatz equations for 
both odd and even $N$ are given by 
\be
{\delta(u_{j}^{(1)})\ h^{(2)}(u_{j}^{(1)}-\eta)
\over \delta(u_{j}^{(1)}-\eta)\ h^{(1)}(u_{j}^{(1)})} 
&=&-{Q_{2}(u_{j}^{(1)}-\eta)\over Q_{2}(u_{j}^{(1)}+\eta)} \,, \qquad j =
1\,, 2\,, \ldots \,, M_{1} \,, \non \\
{h^{(1)}(u_{j}^{(2)}-\eta)\over h^{(2)}(u_{j}^{(2)})}
&=&-{Q_{1}(u_{j}^{(2)}+\eta)\over Q_{1}(u_{j}^{(2)}-\eta)} \,, \qquad j =
1\,, 2\,, \ldots \,, M_{2} \,.
\label{BAEII1}
\ee
where 
\be
\delta(u) &=& 2^{4}\left( \sinh u \sinh(u + 2\eta) \right)^{2N} {\sinh 2u
\sinh (2u + 4\eta)\over \sinh(2u+\eta) \sinh(2u+3\eta)}
\sinh(u+\eta+\alpha_{-})\non \\
& & \sinh(u+\eta-\alpha_{-})\sinh(u+\eta+\alpha_{+})\sinh(u+\eta-\alpha_{+})\cosh^{4}(u + \eta)
\label{delta}
\ee 
and
\be
Q_{a}(u) = \prod_{j=1}^{M_{a}} 
\sinh (u - u_{j}^{(a)}) \sinh (u + u_{j}^{(a)} + \eta) \,, \qquad a = 
1\,, 2\,, 
\label{QII}
\ee
$M_{1}$ and $M_{2}$ are the number of Bethe roots, $u_{j}^{(1)}$ and $u_{j}^{(2)}$ 
(zeros of $Q_{1}(u)$ and $Q_{2}(u)$ respectively). However, $h^{(1)}(u)$ and $h^{(2)}(u)$ 
differ for odd and even values of $N$, as will be noted below. The energy eigenvalues
in terms of the ``shifted'' Bethe roots ${\tilde u_{j}^{(a)}}$ are given by
\be
E={1\over 2} \sinh^{2}\eta 
\sum_{a=1}^{2}\sum_{j=1}^{M_{a}}{1\over 
\sinh (\tilde u_{j}^{(a)} - {\eta\over2})
\sinh (\tilde u_{j}^{(a)} + {\eta\over2})} + {1\over 2}(N-1) \cosh 
\eta \,.
\label{energyIII}
\ee 
where $\tilde u_{j}^{(a)} \equiv u_{j}^{(a)} + {\eta\over2}$.

\subsection{Even $N$}\label{sec:even}

We begin by recalling \cite{MNS2} the structure of roots distribution for this case. 
The Bethe roots $\tilde u_{j}^{(a)}$ for the lowest energy state have the form
\be
\left\{ \begin{array}{c@{\quad : \quad} l}
\mu \lambda_{j}^{(a,1)}                      & j = 1\,, 2\,, \ldots \,, M_{(a,1)} \\
\mu \lambda_{j}^{(a,2)} + {i \pi\over 2} \,, & j = 1\,, 2\,, \ldots \,, M_{(a,2)}
\end{array} \right. \,, \qquad a = 1\,, 2 \,, 
\label{stringhypothesisIIc}
\ee 
where $\lambda_{j}^{(a,b)}$ are real. Here, $M_{(1,1)}= M_{(2,1)} = {N\over 2}$, and $M_{(1,2)}={p+1\over 
2}$, $M_{(2,2)}={p-1\over 2}$.
The $\mu\lambda_{j}^{(a,1)}$ are the zeros of $Q_{a}(u)$ that form real sea (``sea roots'') and $\mu\lambda_{k}^{(a,2)}$ are 
real parts of the ``extra roots'' (also zeros of $Q_{a}(u)$) which are not part of the ``seas''. 
Hence, there are two ``seas'' of real roots. We employ notations similar to the 
one used in \cite{AN}, 
\be
e_{n}(\lambda) =
{\sinh \left(\mu  (\lambda + {i n\over 2}) \right) 
\over \sinh \left( \mu (\lambda - {i n\over 2}) \right) } \,, \qquad
g_{n}(\lambda) = e_{n}(\lambda \pm {i \pi \over 2 \mu})
= {\cosh \left(\mu  (\lambda + {i n\over 2}) \right) 
\over \cosh \left( \mu (\lambda - {i n\over 2}) \right) } \,.
\label{eandgfunctions}
\ee
Rewriting bulk and boundary parameters \cite{AN}, $\eta = i\mu$, 
$\alpha_{\pm} = i\mu a_{\pm}$\footnote{Bethe Ansatz equations written in this and subsequent
sections are true only for suitable values of $a_{\pm}$, namely
${\nu-1 \over 2} < |a_{\pm}| < {\nu+1 \over 2}$ \,, 
\quad $a_{+} a_{-} > 0$\,, where $\nu = p+1$}, where $\mu = {\pi\over{p+1}}$ and taking 
\be
h^{(1)}(u) &=& {8\sinh^{2N+1}(u+2\eta)\cosh^{2}(u+\eta) \cosh(u+2\eta)\over 
\sinh(2u+3\eta)} \,, \quad h^{(2)}(u) = h^{(1)}(-u-2\eta) \,, \label{h1function}
\ee
the Bethe Ansatz equations (\ref{BAEII1}) for the sea roots then take the following form
\cite{MN2,MNS2} 
\be
\lefteqn{e_{1}(\lambda_{j}^{(1,1)})^{2N+1} 
\left[ g_{1}(\lambda_{j}^{(1,1)}) 
e_{1+2a_{-}}(\lambda_{j}^{(1,1)}) e_{1-2a_{-}}(\lambda_{j}^{(1,1)})
e_{1+2a_{+}}(\lambda_{j}^{(1,1)}) e_{1-2a_{+}}(\lambda_{j}^{(1,1)}) \right]^{-1}} 
\label{BAEIIsea1} \\
& & = -\prod_{k=1}^{N/2} \left[
e_{2}(\lambda_{j}^{(1,1)} - \lambda_{k}^{(2,1)}) 
e_{2}(\lambda_{j}^{(1,1)} + \lambda_{k}^{(2,1)})\right] 
 \prod_{k=1}^{(p-1)/2} \left[ 
g_{2}(\lambda_{j}^{(1,1)} - \lambda_{k}^{(2,2)}) 
g_{2}(\lambda_{j}^{(1,1)} + \lambda_{k}^{(2,2)}) \right] \,, \non 
\ee 
and
\be
\lefteqn{e_{1}(\lambda_{j}^{(2,1)})^{2N+1} 
g_{1}(\lambda_{j}^{(2,1)})^{-1}} 
\label{BAEIIsea2} \\
& & = -\prod_{k=1}^{N/2} \left[
e_{2}(\lambda_{j}^{(2,1)} - \lambda_{k}^{(1,1)}) 
e_{2}(\lambda_{j}^{(2,1)} + \lambda_{k}^{(1,1)})\right] 
\prod_{k=1}^{(p+1)/2} \left[  
g_{2}(\lambda_{j}^{(2,1)} - \lambda_{k}^{(1,2)}) 
g_{2}(\lambda_{j}^{(2,1)} + \lambda_{k}^{(1,2)}) \right] \,, \non 
\ee 
respectively, where $j=1\,, \ldots \,, {N\over 2}$. 
The corresponding ground-state counting functions are 
\be
\lefteqn{\h^{(1)}(\lambda) = {1\over 2 \pi}\Big\{ (2N+1) 
q_{1}(\lambda) - r_{1}(\lambda)
- q_{1+2a_{-}}(\lambda) - q_{1-2a_{-}}(\lambda)
- q_{1+2a_{+}}(\lambda) - q_{1-2a_{+}}(\lambda)} \non \\  
& & -\sum_{k=1}^{N/2}\left[ 
q_{2}(\lambda - \lambda_{k}^{(2,1)}) +
q_{2}(\lambda + \lambda_{k}^{(2,1)})\right] 
- \sum_{k=1}^{(p-1)/2} \left[
r_{2}(\lambda - \lambda_{k}^{(2,2)}) +
r_{2}(\lambda + \lambda_{k}^{(2,2)}) \right] \Big\} \,,
\label{h1even}
\ee 
and
\be
\lefteqn{\h^{(2)}(\lambda) = {1\over 2 \pi}\Big\{ (2N+1) 
q_{1}(\lambda) - r_{1}(\lambda)} \non \\
& & -\sum_{k=1}^{N/2}\left[ 
q_{2}(\lambda - \lambda_{k}^{(1,1)}) +
q_{2}(\lambda + \lambda_{k}^{(1,1)})\right] 
- \sum_{k=1}^{(p+1)/2} \left[
r_{2}(\lambda - \lambda_{k}^{(1,2)}) +
r_{2}(\lambda + \lambda_{k}^{(1,2)}) \right] \Big\} \,.
\label{h2even}
\ee 
where $q_{n}(\lambda)$ and $r_{n}(\lambda)$ are odd functions defined
by
\be
q_{n}(\lambda) &=& \pi + i \ln e_{n}(\lambda) 
= 2 \tan^{-1}\left( \cot(n \mu/ 2) \tanh( \mu \lambda) \right)
\,, \non \\
r_{n}(\lambda) &=&  i \ln g_{n}(\lambda) \,.
\label{logfuncts}
\ee 
These counting functions satisfy the following 
\be
\h^{(l)}(\lambda_{j}) = j \,, \qquad j = 1\,,\ldots\,, {N\over 2}
\label{hidentity}
\ee
In (\ref{hidentity}) above, $l = 1\,,2$ 

\subsection{Odd $N$}\label{sec:odd}

In this section, we present an extension of the previous results to include 
solutions for odd $N$ values. The roots distribution is similar to the previous case, 
but now we have $M_{(1,1)}= M_{(2,1)} = {N+1\over 2}$, and $M_{(1,2)} = M_{(2,2)} = {p-1\over 2}$.
Using the following in (\ref{BAEII1}), 
\be
h^{(1)}(u) &=& {\sinh(u-\alpha_{+}+\eta)\sinh(u+\alpha_{+}+\eta)\sinh^{2N+1}(u+2\eta)\cosh^{2}(u+\eta) \cosh(u+2\eta)\over 
\sinh(2u+3\eta)} \,, \non \\
& & h^{(2)}(u) = h^{(1)}(-u-2\eta) \,, \label{h1functionodd}
\ee
we obtain the Bethe Ansatz equations 
\be
\lefteqn{e_{1}(\lambda_{j}^{(1,1)})^{2N+1} 
\left[ g_{1}(\lambda_{j}^{(1,1)}) 
e_{1+2a_{-}}(\lambda_{j}^{(1,1)}) e_{1-2a_{-}}(\lambda_{j}^{(1,1)})
 \right]^{-1}} 
\label{BAEIIsea1} \\
& & = -\prod_{k=1}^{(N+1)/2} \left[
e_{2}(\lambda_{j}^{(1,1)} - \lambda_{k}^{(2,1)}) 
e_{2}(\lambda_{j}^{(1,1)} + \lambda_{k}^{(2,1)})\right] 
 \prod_{k=1}^{(p-1)/2} \left[ 
g_{2}(\lambda_{j}^{(1,1)} - \lambda_{k}^{(2,2)}) 
g_{2}(\lambda_{j}^{(1,1)} + \lambda_{k}^{(2,2)}) \right] \,, \non 
\ee 
and
\be
\lefteqn{e_{1}(\lambda_{j}^{(2,1)})^{2N+1} 
\left[ g_{1}(\lambda_{j}^{(2,1)}) 
e_{1+2a_{+}}(\lambda_{j}^{(2,1)}) e_{1-2a_{+}}(\lambda_{j}^{(2,1)})
 \right]^{-1}} 
\label{BAEIIsea1} \\
& & = -\prod_{k=1}^{(N+1)/2} \left[
e_{2}(\lambda_{j}^{(2,1)} - \lambda_{k}^{(1,1)}) 
e_{2}(\lambda_{j}^{(2,1)} + \lambda_{k}^{(1,1)})\right] 
 \prod_{k=1}^{(p-1)/2} \left[ 
g_{2}(\lambda_{j}^{(2,1)} - \lambda_{k}^{(1,2)}) 
g_{2}(\lambda_{j}^{(2,1)} + \lambda_{k}^{(1,2)}) \right] \,, \non 
\ee 
respectively, where $j=1\,, \ldots \,, {N+1\over 2}$. Note the
presence of parameter-dependant terms in both the equations above.   
One can also notice the number of extra roots changes from ${p+1\over 2}$
to ${p-1\over 2}$ for $Q_{1}(u)$. 
The ground-state counting functions for this case read 
\be
\lefteqn{\h^{(1)}(\lambda) = {1\over 2 \pi}\Big\{ (2N+1) 
q_{1}(\lambda) - r_{1}(\lambda)
- q_{1+2a_{-}}(\lambda) - q_{1-2a_{-}}(\lambda)} \non \\  
& & -\sum_{k=1}^{(N+1)/2}\left[ 
q_{2}(\lambda - \lambda_{k}^{(2,1)}) +
q_{2}(\lambda + \lambda_{k}^{(2,1)})\right] 
- \sum_{k=1}^{(p-1)/2} \left[
r_{2}(\lambda - \lambda_{k}^{(2,2)}) +
r_{2}(\lambda + \lambda_{k}^{(2,2)}) \right] \Big\} \,,
\label{h1odd}
\ee 
and
\be
\lefteqn{\h^{(2)}(\lambda) = {1\over 2 \pi}\Big\{ (2N+1) 
q_{1}(\lambda) - r_{1}(\lambda)
- q_{1+2a_{+}}(\lambda) - q_{1-2a_{+}}(\lambda)} \non \\  
& & -\sum_{k=1}^{(N+1)/2}\left[ 
q_{2}(\lambda - \lambda_{k}^{(1,1)}) +
q_{2}(\lambda + \lambda_{k}^{(1,1)})\right] 
- \sum_{k=1}^{(p-1)/2} \left[
r_{2}(\lambda - \lambda_{k}^{(1,2)}) +
r_{2}(\lambda + \lambda_{k}^{(1,2)}) \right] \Big\} \,,
\label{h2odd}
\ee 
As for even $N$, we again have the following
\be
\h^{(l)}(\lambda_{j}) = j \,, \qquad j = 1\,,\ldots\,,{N+1\over 2}
\label{hidentity2}
\ee
where $l = 1\,,2$. Note that (\ref{hidentity}) and (\ref{hidentity2}) can be
written more compactly as
\be
\h^{(l)}(\lambda_{j}) = j \,, \qquad j = 1\,,\ldots\,,\lfloor{N+1\over 2}\rfloor
\label{hidentity3}
\ee   
where $\lfloor\ldots\rfloor$ denotes the integer part and $\mu\lambda_{\lfloor{N+1\over 2}\rfloor}$ is 
the largest sea root for that ``sea''. Subsequently, we shall denote largest sea roots as 
$\mu\Lambda_{l}$.

\section{Finite-size correction of order $1/N$}\label{sec:finitesize}

In this section, we shall compute the finite-size correction for the ground state and low lying excited states.
For these excited states, we restrict our analysis to excitations by holes
which are located to the right of the real sea roots. Applying (\ref{stringhypothesisIIc}) 
to (\ref{energyIII}), we get the lowest state energy eigenvalues for chain of finite length
$N$, 
\be
E &=& - {\pi \sin \mu\over \mu} \Big\{ 
{1\over 2}\sum_{a=1}^{2}\sum_{j=-\lfloor{N+1\over2}\rfloor}^{\lfloor{N+1\over2}\rfloor} 
a_{1}(\lambda_{j}^{(a,1)}) -a_{1}(0)
+ \sum_{a=1}^{2}\sum_{j=1}^{M_{(a,2)}} b_{1}(\lambda_{j}^{(a,2)} ) \Big\} \non \\ 
& & + {1\over 2}(N-1) \cos \mu  \,. \label{energyII2} 
\ee 
where notations from \cite{AN} have again been adopted
\be
a_n(\lambda) &=& {1\over 2\pi} {d \over d\lambda} q_n (\lambda)
= {\mu \over \pi} 
{\sin (n \mu)\over \cosh(2 \mu \lambda) - \cos (n \mu)} \,, \non \\
b_n(\lambda) &=& {1\over 2\pi} {d \over d\lambda} r_n (\lambda)
= -{\mu \over \pi} 
{\sin (n \mu)\over \cosh(2 \mu \lambda) + \cos (n \mu)} \,. 
\label{anbn}
\ee 
\noindent Note that $M_{(a,2)}$ in (\ref{energyII2}), refers to number of extra
roots for $Q_{a}(u)$. The first and third terms in the curly bracket of (\ref{energyII2})
are summed over the number of sea roots and extra roots respectively. As one 
considers next lowest excited state, the number of sea roots and extra roots change. 
Hence, for these states of low lying excitations (with real sea), the very same 
term in the first sum will again be summed over accordingly between approriate 
limits dictated by the number of sea roots. As for the summation over extra roots,
the function summed over depends on the imaginary part of these roots, especially
in the presence of 2-strings. However, as one shall see, for $1/N$ correction 
(in the $N\rightarrow\infty$ limit), only the sum over the sea roots contributes.
The second sum in (\ref{energyII2}) contributes to order 1 correction (boundary energy)
which we have considered elsewhere \footnote{Equation [4.26] for the boundary 
energy in \cite{MNS2} holds both for even and odd values of $N$}\cite{MNS2}.

\subsection{Sum-rule and hole-excitations }\label{sec:sumrule}
 
Now we present some results based on the solution of the model (\ref{Hamiltonian})
for $N = 2\,,3\,,\ldots\,,7$. We begin with even $N$ case. We find for even 
$N$, excited states contain odd number of holes for each $Q_{a}(u)$. This can be 
seen from the following analysis on counting functions.
For the lowest energy state the counting functions are given by (\ref{h1even}) and (\ref{h2even}).
By using the fact that $q_{n}(\lambda)\rightarrow$ sgn$(n)\pi - \mu n$ and 
$r_{n}(\lambda)\rightarrow -\mu n$ as $\lambda\rightarrow\infty$ and $\rho^{(l)} = {1\over N}{d\h^{(l)}\over d\lambda}$ 
we have the following sum rule
\be
\int_{\Lambda_{l}}^{\infty}d\lambda\ \rho^{(l)}(\lambda) &=& {1\over N}\big(\h^{(l)}(\infty)-\h^{(l)}(\Lambda_{l})\big) \non \\
&=& {1\over N}({1\over 2} + 1)
\label{sumrulelowesteven}
\ee
$\mu\Lambda_{l}$ refers to the largest sea root. As before $l = 1\,,2$.  
We make use of the fact that 
\be
\h^{(l)}(\infty) = {N\over 2} + {3\over 2} \non \\ 
\h^{(l)}(\Lambda_{l}) = {N\over 2}
\label{limit1}
\ee
From (\ref{sumrulelowesteven}) and (\ref{limit1}), we see that there is one hole located to the right of the largest sea root.
Similar analysis for low lying (multi-hole) excited states yields the following 
\be
\int_{\Lambda_{l}}^{\infty}d\lambda\ \rho^{(l)}(\lambda) &=& {1\over N}\big(\h^{(l)}(\infty)-\h^{(l)}(\Lambda_{l})\big) \non \\
&=& {1\over N}({1\over 2} + N_{H})
\label{sumruleexcitedeven}
\ee
where $N_{H}$ is the number of holes (odd) to the right of the corresponding largest sea root.
To illustrate the results above, we consider the following low lying excited states 
with ${N\over 2}-1$ and ${N\over 2}-2$ sea roots and therefore
different number of extra roots than the lowest energy state \footnote{The lowest energy state has ${N\over 2}$
sea roots. As for the extra roots, there are ${p+1\over 2}$ and ${p-1\over 2}$ of them for $Q_{1}(u)$ and $Q_{2}(u)$ respectively}.
The former case is found to have one hole with ${p-1\over 2}$ and ${p-3\over 2}$
extra roots in addition to a 2-string from each of the $Q_{1}(u)$ and $Q_{2}(u)$
respectively. From,
\be   
\h^{(l)}(\infty) = {N\over 2} + {1\over 2} \non \\ 
\h^{(l)}(\Lambda_{l}) = {N\over 2}-1
\label{limit2}
\ee
one has 
\be
{1\over N}\big(\h^{(l)}(\infty)-\h^{(l)}(\Lambda_{l})\big) = {1\over N}({1\over 2} + 1)
\label{sumruleexcitedevencaseI}
\ee
Hence giving $N_{H} = 1$. The later case has three holes with ${p+1\over 2}$ and ${p-1\over 2}$ extra roots 
and a 2-string from each of the $Q_{a}(u)$ with $a = 1\,,2$. Similar analysis,
\be
\h^{(l)}(\infty) = {N\over 2} + {3\over 2} \non \\ 
\h^{(l)}(\Lambda_{l}) = {N\over 2}-2
\label{limit3}
\ee
yields
\be
{1\over N}\big(\h^{(l)}(\infty)-\h^{(l)}(\Lambda_{l})\big) = {1\over N}({1\over 2} + 3)
\label{sumruleexcitedevencaseII}
\ee
giving $N_{H} = 3$. The total number of roots are the same for all these states. 
There are also excited states with equal number of sea and extra roots as for the
state of lowest energy, but with position of the single hole nearer to the origin than that of the 
lowest energy state, suggesting the usual bulk hole-excitation scenario, $E_{hole}(\lambda^{(a)})$
increases as $\lambda^{(a)}\rightarrow 0$ where $E_{hole}(\lambda^{(a)})$ is the energy
due to the presence of holes and $\lambda^{(a)}$, with $a = 1\,,2$ denote the positions of the holes in both ``seas''. 
We shall compute the explicit expression for energy due to holes shortly. 

As for the odd $N$ case, we have the true ground state, namely state 
of lowest energy without hole. From the counting functions, 
(\ref{h1odd}) and (\ref{h2odd}), we have
\be
\int_{\Lambda_{l}}^{\infty}d\lambda\ \rho^{(l)}(\lambda) &=& {1\over N}\big(\h^{(l)}(\infty)-\h^{(l)}(\Lambda_{l})\big) \non \\
&=& {1\over 2 N}
\label{sumrulelowestodd}
\ee
As before $l = 1\,,2$, and we make use of the fact that 
\be
\h^{(l)}(\infty) = {N\over 2} + 1 \non \\ 
\h^{(l)}(\Lambda_{l}) = {N+1\over 2}
\label{limitodd1}
\ee
From (\ref{limitodd1}), we see that this state of lowest energy for odd $N$ 
has no hole, signifying the true ground state. Similar analysis for low lying 
excited states yields the following 
\be
\int_{\Lambda_{l}}^{\infty}d\lambda\ \rho^{(l)}(\lambda) &=& {1\over N}\big(\h^{(l)}(\infty)-\h^{(l)}(\Lambda_{l})\big) \non \\
&=& {1\over N}({1\over 2} + N_{H})
\label{sumruleexcitedodd}
\ee
where $N_{H}$ is the number of holes (even) to the right of sea roots. 
Hence, for odd $N$ case, there are even number of holes (for each $Q_{a}(u)$),
with $a = 1\,,2$, for the excited states, e.g., for the first excited state with ${N-1\over 2}$ sea roots,  
\be
\h^{(l)}(\infty) = {N+1\over 2} + {3\over 2} \non \\ 
\h^{(l)}(\Lambda_{l}) = {N-1\over 2}
\label{limit3odd}
\ee
which signifies the presence of two holes.

It is known for simpler models of spin chains e.g., 
closed XXZ chain that even number of holes are present in chains with 
even number of spins and vice versa. Hence, the true ground state (lowest energy
state with no holes) for these models is found to lie in even $N$ sector. 
The reverse scenario (one hole in the lowest energy state for even $N$ and ground state in odd $N$ sector) 
we find here for this model can be explained using some heuristic arguments based on spin and magnetic fields at the two 
boundaries, similar to the one given in Section 3 of \cite{MNS2} \footnote{Readers
are urged to refer to Figures 2 and 3 in that Section}. In footnote 2, 
we notice the signs of $a_{+}$ and $a_{-}$ must be the same for boundary parameter region of interest.
Hence, in Hamiltonian (\ref{Hamiltonian}), the direction of
the magnetic fields at the two boundaries are also the same (Both up or both down).
This upsets the antiferromagnetic spin arrangement at the boundaries,
favouring spin allignments along the same direction at the boundaries for chains with even $N$.
This causes the following: presence of odd $N$ behaviours in the even $N$ chain, namely the lowest
energy state for even $N$ sector has one hole for each $Q_{a}(u)$. Spins at the boundaries for the odd
$N$ chain will not experience such spin upset since the parallel magnetic fields
favours the antiferromagnetic arrangement of an odd $N$ chain. Therefore, the lowest
energy state for odd $N$ chain has no holes. In other words, the true ground state 
exists in odd $N$ sector. Further effects are the presence of odd and even
number of holes in chains with even and odd $N$ respectively as shown in the analysis above.

Now, the energy due to hole excitations can be presented. We consider 
first the lowest energy state for even $N$ case with one hole. Using 
\be
{1\over N}\sum_{k = -{N\over 2}}^{{N\over 2}}g(\lambda-\lambda_{k}^{(a,1)})\approx
\int_{-\infty}^{\infty}d\lambda'\ \rho^{(l)}(\lambda')g(\lambda-\lambda') - {1\over N}
g(\lambda-\tilde{\lambda}^{(a)})
\label{approx}
\ee
for some arbitrary function $g(\lambda)$ and
\be
\rho^{(l)} = {1\over N}{d\h^{(l)}\over d\lambda}
\label{density}
\ee
where $l= 1\,,2$, $\mu\lambda_{k}^{(a,1)}\equiv$ sea roots, with $a = 1\,,2$, 
and $\mu\tilde{\lambda}^{(a)}\equiv$ position of the hole for each of the $Q_{a}(u)$, 
one can write down the sum of the two densities 
\be
\rho^{(1)}(\lambda) + \rho^{(2)}(\lambda) &=& 4 a_{1}(\lambda)-\int_{-\infty}^{\infty}d\lambda'\ (\rho^{(1)}(\lambda') + \rho^{(2)}(\lambda'))a_{2}(\lambda-\lambda')\non \\
&+& {1\over N}\big[a_{2}(\lambda-\tilde{\lambda}^{(1)}) + a_{2}(\lambda-\tilde{\lambda}^{(2)})\big]
+ {1\over N}\big[2 a_{1}(\lambda) + 2 a_{2}(\lambda) - 2 b_{1}(\lambda)\non \\
&-& a_{1+2a_{-}}(\lambda) - a_{1-2a_{-}}(\lambda) - a_{1+2a_{+}}(\lambda) - a_{1-2a_{+}}(\lambda)\non \\ 
&-& \sum_{k=1}^{{p-1\over 2}}(b_{2}(\lambda-\lambda_{k}^{(2,2)}) 
+ b_{2}(\lambda + \lambda_{k}^{(2,2)}))\non \\
&-& \sum_{k=1}^{{p+1\over 2}}(b_{2}(\lambda-\lambda_{k}^{(1,2)}) + b_{2}(\lambda+\lambda_{k}^{(1,2)}))\big]
\label{sumofdensities}
\ee  
Defining $\rho_{total}(\lambda)\equiv\rho^{(1)}(\lambda) + \rho^{(2)}(\lambda)$ 
and solving (\ref{sumofdensities}) using Fourier transform \footnote{Our 
conventions are
\be
\hat f(\omega) \equiv \int_{-\infty}^\infty e^{i \omega \lambda}\ 
f(\lambda)\ d\lambda \,, \qquad\qquad
f(\lambda) = {1\over 2\pi} \int_{-\infty}^\infty e^{-i \omega \lambda}\ 
\hat f(\omega)\ d\omega \,. \non 
\ee}, we have
\be
\hat \rho_{total}(\omega) &=& 4 \hat s(\omega) + {1\over N}\hat R(\omega)\non \\
&+& {1\over N}\hat J(\omega)\big(e^{i\omega \tilde {\lambda}^{(1)}} +
e^{i\omega \tilde {\lambda}^{(2)}}\big)   
\label{solutionfordensity}
\ee
where $\hat \rho_{total}(\omega)\,,\hat a_{2}(\omega)$ \footnote{\be
\hat a_{n}(\omega) &=& \sgn(n) {\sinh \left( (\nu  - |n|) 
\omega / 2 \right) \over
\sinh \left( \nu \omega / 2 \right)} \,,
\qquad 0 \le |n| < 2 \nu   \non \label{fourier1}. \non  
\label{fourier2}
\ee} and $\hat s(\omega)$ 
are the Fourier transforms of $\rho_{total(\lambda)}\,,a_{2}(\lambda)$ and
${a_{1}(\lambda)\over 1 + a_{2}(\lambda)}$ respectively. Also $\hat J(\omega) = {\hat a_{2}(\omega)\over 1 + \hat a_{2}(\omega)}$.
$\hat R(\omega)$ is the contribution from the second square bracket in (\ref{sumofdensities}), which
will not enter the calculation for $E_{hole}(\tilde{\lambda}^{(a)})$ and will be omitted henceforth. The 
Fourier transform of hole density are the third and the fourth terms in (\ref{solutionfordensity}), which gives
\be
\rho_{hole}(\lambda) = {1\over N}\big[J(\lambda-\tilde {\lambda}^{(1)})+
J(\lambda-\tilde {\lambda}^{(2)})\big]
\label{holedensity}
\ee
Using approximation (\ref{approx}) in (\ref{energyII2}), and making use of (\ref{holedensity}), one has
\be
E_{hole}(\tilde{\lambda}^{(a)}) &=& -{N\pi\sin\mu\over 2 \mu}\int_{-\infty}^{\infty}d\lambda\ a_{1}(\lambda)\rho_{hole}(\lambda)\non \\
&+& {\pi\sin\mu\over 2 \mu} \sum_{a=1}^{2}a_{1}(\tilde{\lambda}^{(a)})
\label{energyhole1}
\ee
which after some manipulation yields
\be
E_{hole}(\tilde{\lambda}^{(a)}) &=& {\pi\sin\mu\over 4 \mu}\sum_{a=1}^{2}{1\over \cosh\pi\tilde{\lambda}^{(a)}}
\label{energyhole2}
\ee
Generalizing the derivation to $\alpha$ number of holes, one has
\be
\rho_{hole}(\lambda) = {1\over N}\sum_{\alpha}\sum_{a=1}^{2}J(\lambda-\tilde {\lambda}_{\alpha}^{(a)})
\label{holedensitygeneral}
\ee
and finally the following for the energy
\be
E_{hole}(\tilde{\lambda}_{\alpha}^{(a)}) &=& {\pi\sin\mu\over 4 \mu}\sum_{\alpha}\sum_{a=1}^{2}{1\over \cosh\pi\tilde{\lambda}_{\alpha}^{(a)}}
\label{energyhole2general}
\ee
Note that $E_{hole}(\tilde{\lambda}_{\alpha}^{(a)})$ increases as $\tilde{\lambda}_{\alpha}^{(a)}\rightarrow 0$ 
as mentioned above in paragraph following (\ref{sumruleexcitedevencaseII}).

\subsection{Casimir energy}\label{sec:casimir}

In this section, we give the derivation of $1/N$ correction (Casimir energy)
to the lowest energy state, for the even $N$ case (with one hole). This result 
is then generalized to include odd $N$ values as well as the low lying (multi-hole) excited 
states. We begin by presenting the expression for the density difference between
chain of finite length (with $N$ spins), $\rho_{N}^{(1)}(\lambda)+\rho_{N}^{(2)}(\lambda)$ 
and that of infinite length, $\rho_{\infty}(\lambda)$
\be
\rho_{N}^{(1)}(\lambda)+\rho_{N}^{(2)}(\lambda) - \rho_{\infty}(\lambda) &=&
- \int_{-\infty}^{\infty}d\gamma\ a_{2}(\lambda-\gamma)\big[{1\over N}\sum_{\beta = -{N\over 2}}^{{N\over 2}}
\delta(\gamma-\lambda_{\beta}^{(1,1)})-\rho_{N}^{(1)}(\gamma)\big] \non \\
&-& \int_{-\infty}^{\infty}d\gamma\ a_{2}(\lambda-\gamma)\big[{1\over N}\sum_{\beta = -{N\over 2}}^{{N\over 2}}
\delta(\gamma-\lambda_{\beta}^{(2,1)})-\rho_{N}^{(2)}(\gamma)\big]\non \\
&-& \int_{-\infty}^{\infty}d\gamma\ a_{2}(\lambda-\gamma)\big[\rho_{N}^{(1)}(\gamma)+\rho_{N}^{(2)}(\gamma)-\rho_{\infty}(\gamma)\big]
\label{densitydiff}
\ee
In (\ref{densitydiff}) and henceforth, only terms that are crucial to the
computation of $1/N$ correction are given. Other parameter dependant terms that
contribute to order 1 correction have been omitted here
\footnote{See \cite{MNS2} for details}. Solving (\ref{densitydiff}) yields
\be
\rho_{N}^{(1)}(\lambda)+\rho_{N}^{(2)}(\lambda) - \rho_{\infty}(\lambda) &=&
- \int_{-\infty}^{\infty}d\gamma\ p(\lambda-\gamma)\big[{1\over N}\sum_{\beta = -{N\over 2}}^{{N\over 2}}
\delta(\gamma-\lambda_{\beta}^{(1,1)})-\rho_{N}^{(1)}(\gamma)\big] \non \\
&-& \int_{-\infty}^{\infty}d\gamma\ p(\lambda-\gamma)\big[{1\over N}\sum_{\beta = -{N\over 2}}^{{N\over 2}}
\delta(\gamma-\lambda_{\beta}^{(2,1)})-\rho_{N}^{(2)}(\gamma)\big]
\label{densitydiff2}
\ee
where $\rho_{\infty}(\lambda) = {4 a_{1}(\lambda)\over 1 + a_{2}(\lambda)} \equiv 4 s(\lambda)$ 
and $p(\lambda) = {1\over 2\pi}\int_{-\infty}^{\infty}d\omega\ e^{-i\omega\lambda}{\hat a_{2}(\omega)\over 1 + \hat a_{2}(\omega)}$
Similar equation expressing the energy difference between finite and infinite system
is also needed to compute Casimir energy. This is given by
\be
E_{N}-E_{\infty} &=& -{N\pi \sin\mu\over 2\mu}\Big\{\int_{-\infty}^{\infty}d\lambda\ a_{1}(\lambda)\big[{1\over N}\sum_{\beta = -{N\over 2}}^{{N\over 2}}
\delta(\lambda-\lambda_{\beta}^{(1,1)})-\rho_{N}^{(1)}(\lambda)\big] \non \\
&+& \int_{-\infty}^{\infty}d\lambda\ a_{1}(\lambda)\big[{1\over N}\sum_{\beta = -{N\over 2}}^{{N\over 2}}
\delta(\lambda-\lambda_{\beta}^{(2,1)})-\rho_{N}^{(2)}(\lambda)\big]\non \\
&+& \int_{-\infty}^{\infty}d\lambda\ a_{1}(\lambda)\big[\rho_{N}^{(1)}(\lambda)+\rho_{N}^{(2)}(\lambda)-\rho_{\infty}(\lambda)\big]\Big\} 
\label{energydiff1}
\ee 
Using (\ref{densitydiff2}) and the fact that $\hat p(\omega) \hat a_{1}(\omega) = \hat s(\omega) \hat a_{2}(\omega)$,
we have
\be
E_{N}-E_{\infty} &=& -{N\pi \sin\mu\over 4\mu}\Big\{\int_{-\infty}^{\infty}d\lambda\
S_{N}^{(1)}(\lambda)\rho_{\infty}^{(1)}(\lambda) + \int_{-\infty}^{\infty}d\lambda\
S_{N}^{(2)}(\lambda)\rho_{\infty}^{(2)}(\lambda)\Big\}
\label{energydiff2}
\ee 
where
$S_{N}^{(l)}(\lambda)\equiv {1\over N}\sum_{\beta = -{N\over 2}}^{N\over 2}\delta(\lambda - \lambda_{\beta}^{(l,1)}) - \rho_{N}^{(l)}(\lambda)$
and $\rho_{\infty}^{(l)}(\lambda) = {1\over 2}\rho_{\infty}(\lambda)\equiv 2 s(\lambda)$ with $l = 1\,,2$.
Further, using Euler-Maclaurin summation formula \cite{WW}, (\ref{energydiff2}) becomes
\be
E_{N}-E_{\infty} &=& -{N\pi \sin\mu\over 2\mu}\Big\{-\int_{\Lambda_{1}}^{\infty}d\lambda\ \rho_{\infty}^{(1)}(\lambda)\rho_{N}^{(1)}(\lambda)
+ {1\over 2 N}\rho_{\infty}^{(1)}(\Lambda_{1}) + {1\over 12 N^{2}\rho_{N}^{(1)}(\Lambda_{1})}\rho_{\infty}^{(1)'}(\Lambda_{1})\non \\ 
&-& \int_{\Lambda_{2}}^{\infty}d\lambda\ \rho_{\infty}^{(2)}(\lambda)\rho_{N}^{(2)}(\lambda)
+ {1\over 2 N}\rho_{\infty}^{(2)}(\Lambda_{2}) + {1\over 12 N^{2}\rho_{N}^{(2)}(\Lambda_{2})}\rho_{\infty}^{(2)'}(\Lambda_{2})\Big\}
\label{EMac}
\ee
(\ref{densitydiff2}) can also be expressed in similar form
\be
\rho_{N}^{(1)}(\lambda)+\rho_{N}^{(2)}(\lambda)&-&\rho_{\infty}(\lambda) = 
\int_{\Lambda_{1}}^{\infty}d\gamma\ p(\lambda-\gamma) \rho_{N}^{(1)}(\gamma) -
{1\over 2N}p(\lambda-\Lambda_{1}) - {p^{'}(\lambda-\Lambda_{1})\over 12 N^{2}
\rho_{N}^{(1)}(\Lambda_{1})}\non \\
&+&\int_{\Lambda_{2}}^{\infty}d\gamma\ p(\lambda-\gamma) \rho_{N}^{(2)}(\gamma) - 
{1\over 2N}p(\lambda-\Lambda_{2}) - {p^{'}(\lambda-\Lambda_{2})\over 12 N^{2}
\rho_{N}^{(2)}(\Lambda_{2})} 
\label{EMacfordensity}
\ee
As before, $\mu\Lambda_{1}$ and $\mu\Lambda_{2}$ are the largest sea roots from the two ``seas'' respectively. 
From this point, the calculation very closely resembles the details found 
in Section 2 in \cite{Hamer}. Hence, we omit the details and give only the crucial steps.
Note that (\ref{EMacfordensity}) can be written in the standard form of the
Wiener-Hopf equation \cite{MF} after redefining the terms,
\be
\chi^{(1)}(t) + \chi^{(2)}(t)&-&\int_{0}^{\infty}ds\ p(t-s)\chi^{(1)}(s) - \int_{0}^{\infty}ds\ p(t-s)\chi^{(2)}(s) \non \\
&\approx& f^{(1)}(t) - {1\over 2 N}p(t) + {1\over 12 N^{2}\rho_{N}^{(1)}(\Lambda_{1})}p^{'}(t)\non \\ 
&+& f^{(2)}(t) - {1\over 2 N}p(t) + {1\over 12 N^{2}\rho_{N}^{(2)}(\Lambda_{2})}p^{'}(t)  
\label{WH}
\ee
where the following definitions have been adopted
\be
\chi^{(l)}(\lambda) = \rho_{N}^{(l)}(\lambda + \Lambda_{l})\non \\
f^{(l)}(\lambda) = \rho_{\infty}^{(l)}(\lambda + \Lambda_{l})
\label{shifts}
\ee
and following change in variable is used : $t = \lambda - \Lambda_{l}$ with $l = 1\,,2$
From the Fourier transformed version of (\ref{WH}), one can solve for $X_{+}^{(l)}(\omega)$ 
which is the Fourier transfrom of $\chi_{+}^{(l)}(t)$ 
that is analytic in the upper half complex plane \footnote{Again for complete
details, refer to \cite{Hamer}},
\be
\hat X_{+}^{(l)}(\omega) &=& {1\over 2 N} + {i\omega\over 12 N^{2}\rho_{N}^{(l)}(\Lambda_{l})}\non \\
&+& G_{+}(\omega)\Big[{ig_{1}\over 12 N^{2}\rho_{N}^{(l)}(\Lambda_{l})}-{1\over 2 N} - {i\omega\over 12 N^{2}\rho_{N}^{(l)}(\Lambda_{l})}\non \\ 
&+& {\pi\over \pi-i\omega}\Big({\alpha\over N}+{1\over 2N}-{ig_{1}\over 12 N^{2}\rho_{N}^{(l)}(\Lambda_{l})}\Big)\Big]
\label{Xomega}
\ee 
where $G_{+}(\omega)G_{+}(-\omega) = 1 + \hat a_{2}(\omega)$, $g_{1} = {i\over 12}\big(2 + \nu - {2\nu\over \nu-1}\big)$ and
$\alpha = {1\over G_{+}(0)} = ({\nu\over 2 (\nu -1)})^{1\over 2}$, with
$G_{+}(0)^{2}={2(\nu-1)\over \nu}$. 
\newline From (\ref{sumrulelowesteven}), (\ref{shifts}) and (\ref{Xomega}), one can then determine $\rho_{N}^{(1)}(\Lambda_{1})$ and
$\rho_{N}^{(2)}(\Lambda_{2})$ explicitly from
\be
\chi_{+}^{(l)}(0)\equiv {1\over 2}\rho_{N}^{(l)}(\Lambda_{l}) = {1\over 2 \pi}\int_{-\infty}^{\infty}d\omega\ \hat X_{+}^{(l)}(\omega)
\label{contintegrate}
\ee
by contour integration and some algebra. We give the result below
\be
\rho_{N}^{(l)}(\Lambda_{l}) = {1\over 4 N}\Big\{\pi + 2\pi\alpha + ig_{1} + \big[\pi^{2}
+ {2ig_{1}\pi\over 3} - {g_{1}^2\over 3} + 4\pi^{2}\alpha^{2} + 4\pi\alpha (\pi + ig_{1})\big]^{{1\over 2}}\Big\}
\label{densityatlargestroot}
\ee  
\newline Finally, using $\rho_{\infty}^{(l)}(\lambda)\approx 2 e^{-\pi\lambda}$ for $\lambda\rightarrow \Lambda_{l}$ and (\ref{EMac}), one arrives 
at the desired expression for $1/N$ correction to the energy,
\be
E_{N}-E_{\infty} = E_{Casimir} = -{\pi^{2}\sin\mu\over 24 \mu N}(1-12\alpha^{2})
\label{Casimirenergy}
\ee
where the effective central charge is
\be
c_{eff} &=& 1-12\alpha^{2}\non \\
&=& 1-6{\nu\over (\nu - 1)}
\label{centralcharge}
\ee
We see that for this model, the central charge, $c = 1$ (Free boson). Also $c_{eff}$ is 
independent of boundary parameters, unlike for the Dirichlet case \cite{Hamer}.
This is a feature expected for models with Neumann boundary condition. 
Further, from conformal field theory, one also has the following for the conformal dimensions, 
\be
\Delta &=& {1 - c_{eff}\over 24}\non \\
&=& {\nu\over 4 (\nu - 1)}
\label{conformaldim}
\ee
Note that the above results are derived for the lowest energy state for even $N$ with one
hole for each $Q_{a}(u)$. Reviewing the derivation above, one can notice
that the results above can be further generalized for any $N$ and for low lying excited states
with arbitrary number of holes, provided these holes are located to the right
of the largest sea root as mentioned in the beginning of Section \ref{sec:finitesize}. 
For these excited states, the sum for $S_{N}^{(l)}(\lambda)$ in (\ref{densitydiff}) - 
(\ref{energydiff2}) will inevitably have different limits since the number of 
sea roots vary. However, after applying the Euler-Maclaurin formula, one would
recover (\ref{EMac}) and (\ref{EMacfordensity}). In addition to that, for states with
$N_{H}$ number of holes (all located to the right of the largest sea root), one uses the 
more general result for the sum rule, namely (\ref{sumruleexcitedeven})
and ({\ref{sumruleexcitedodd}) which eventually yields 
\be
\alpha = {N_{H}\over G_{+}(0)}
\label{generalalpha}
\ee
Thus, we have the following for the effective central charge and conformal dimensions
for low lying excited states
\be
c_{eff} &=& 1-6{\nu\over (\nu - 1)}N_{H}^{2}\non \\
\Delta &=& {\nu\over 4 (\nu - 1)}N_{H}^{2}
\label{centralchargegeneral}
\ee
Surprisingly, the results (\ref{conformaldim}) and (\ref{centralchargegeneral}) appear 
to have more resemblance to spin chains with diagonal boundary terms, as one could see
from the ${\nu\over \nu - 1}$ dependance \cite{LeClair}-\cite{Affleck2}, rather than 
${\nu -1\over \nu}$ \cite{AhnZoltan} which is the anticipated form for conformal dimensions for spin chains with
nondiagonal boundary terms. Indeed the theory of a free Bosonic field $\varphi$ 
compactified on a circle of radius $r$ is invariant under $\varphi\mapsto \varphi + 2\pi r$, 
where $r = {2\over\beta}$. $\beta$ is the continuum bulk coupling constant that 
is related to $\nu$ by $\beta^{2} = 8\pi\big({\nu - 1\over\nu }\big)$. Further, the quantization of 
the momentum zero-mode $\Pi_{0}$, yields $\Pi_{0} = {n\beta\over 2}$ for Neumann boundary condition
and $\Pi_{0} = {2n\over \beta}$ for the Dirichlet case, where $n$ is an integer. 
Hence, the zero-mode contribution to the energy, $E_{0,n}\sim \Pi_{0}^{2}$ implies 
$E_{0,n}\sim \Delta\sim \big({\nu-1\over\nu}\big)$ for Neumann and $E_{0,n}\sim \Delta\sim \big({\nu\over\nu-1}\big)$ 
for Dirichlet case respectively. More complete discussion on this topic can be found in 
\cite{Saleur2,AhnZoltan}. Next, we will resort to numerical analysis  
to confirm our analytical results obtained in this section.  
 
\section{Numerical results}\label{sec:numerical}

We present here some numerical results for both odd and even $N$ cases, 
to support our analytical derivations in Section \ref{sec:casimir}. We first solve numerically the 
Bethe equations (\ref{BAEII1}), (\ref{h1even}), (\ref{h2even}), (\ref{h1odd}) and
(\ref{h2odd}) for some large number of spins.  We use these solutions 
to calculate Casimir energy numerically from the following 
\be
E = E_{bulk} + E_{boundary} + E_{Casimir}
\label{casimirboundarybulk}
\ee
In (\ref{casimirboundarybulk}), $E$ is given by (\ref{energyII2}). Thus, having
determined the Bethe roots numerically, one uses known expressions for $E_{bulk}$ \cite{Yang} and
$E_{boundary}$ \cite{MNS2} to determine $E_{Casimir}$. Then using the expression
found above for $E_{Casimir}$, namely (\ref{Casimirenergy}), one can determine the effective central charge, $c_{eff}$ 
for that value of $N$,
\be
c_{eff} = -{24\mu N\over \pi^{2} \sin\mu}(E - E_{bulk} - E_{boundary})
\label{centralchargenumerical}
\ee
Finally, we employ an algorithm due to Vanden Broeck 
and Schwartz \cite{VBS}--\cite{Hamer3} to extrapolate these values for central charge at $N\rightarrow \infty$ limit.  
Table 1 below shows the $c_{eff}$ values for some finite even $N$, for the lowest energy state
with one hole ($N_{H} = 1$). Equation (\ref{centralchargegeneral}) predicts $c_{eff}$
values of -11 and -7 for $p = 1$ and $p = 3$ \footnote{$\nu = p+1$} respectively which are the extrapolated
values (-11.000315 and -7.000410) we obtain from the Vanden Broeck and Schwartz method.

\begin{table}[htb] 
  \centering
  \begin{tabular}{|c|c|c|}\hline
    $N$ & $c_{eff}$, $p = 1\,,\nu = 2$  & $c_{eff}$, $p = 3\,,\nu = 4$\\
    \hline
    16      & -9.365620       & -2.853872 \\ 
    24      & -9.857713       & -3.271279 \\
    32      & -10.122128      & -3.557148 \\
    40      & -10.287160      & -3.770882 \\
    48      & -10.399970      & -3.939554 \\
    56      & -10.481956      & -4.077652 \\ 
    64      & -10.544233      & -4.193784 \\  
    \vdots  &   \vdots        &  \vdots   \\
    $\infty$& -11.000315      & -7.000410 \\ 
    \hline
   \end{tabular}
   \caption{Central charge values, $c_{eff}$ for $p = 1$ ($a_{+}=0.783$, $a_{-}= 0.859$) 
    and $p = 3$ ($a_{+}=2.29$, $a_{-}= 1.76$), 
    from numerical computations based on $N = 16$\,,$24$\,,\ldots\,,$64$ and 
    extrapolated values at $N\rightarrow \infty$ limit (Vanden Broeck and Schwartz algorithm).}
  \label{c2table}
\end{table}

For odd $N$ sector, since $N_{H} = 0$, (\ref{centralchargegeneral}) predicts 
$c_{eff} = 1$ (for the ground state) for any odd $p$. We present
similar numerical results for odd $N$ in Table 2 below for $p = 1$ and
$p = 3$. We work out the $c_{eff}$ values numerically for $N = 15$\,,$25$\,,\ldots\,,$65$.
Excellent agreement between the calculated and the extrapolated values of 1.000770 and 1.001851 
again strongly supports our analytical results.

\begin{table}[htb] 
  \centering
  \begin{tabular}{|c|c|c|}\hline
    $N$ & $c_{eff}$, $p = 1\,,\nu = 2$  & $c_{eff}$, $p = 3\,,\nu = 4$\\
    \hline
    15      & 0.898334      & 0.531501 \\ 
    25      & 0.936128      & 0.634012 \\
    35      & 0.953433      & 0.692758 \\
    45      & 0.963360      & 0.731841 \\
    55      & 0.969797      & 0.760142 \\
    65      & 0.974311      & 0.781795 \\ 
    \vdots  &  \vdots       &  \vdots  \\
    $\infty$& 1.000770      & 1.001851 \\ 
    \hline
   \end{tabular}
   \caption{Central charge values, $c_{eff}$ for $p = 1$ ($a_{+}=0.926$, $a_{-}= 0.654$) 
    and $p = 3$ ($a_{+}=2.10$, $a_{-}= 1.80$), 
    from numerical computations based on $N = 15$\,,$25$\,,\ldots\,,$65$ and 
    extrapolated values at $N\rightarrow \infty$ limit (Vanden Broeck and Schwartz algorithm).}
  \label{c3table}
\end{table}

\section{Discussion}

From the proposed Bethe ansatz equations for an open XXZ spin chain with special nondiagonal boundary terms at roots of unity, we 
computed finite size effect, namely the $1/N$ correction (Casimir energy) to 
the lowest energy state for both even and odd $N$. We also studied the bulk
excitations due to holes. We found some peculiar results for these excitations 
of this model. Firstly, the number of holes for excited states 
seem to be reversed: even number of holes for chains with odd number of spins and vice versa. However,
one could explain this by resorting to heuristic arguments involving effects of magnetic
fields on the spins at the boundary. We then computed the energy due to hole-excitations.
We further generalized the finite-size correction calculation to include multi-hole excited 
states, where these holes are situated to the right of the largest sea root. 
Having found the correction, we proceeded to compute the effective central charge, 
$c_{eff}$ and the conformal dimensions, $\Delta$ for the model. 
We found the central charge, $c = 1$. The effective central charge is independent 
of the boundary parameters, as expected for models with Neumann boundary condition. 
The result for $\Delta$ however, turns out to be similar to models with diagonal boundary terms
rather than the nondiagonal ones, to which the model studied here belongs to.   

As an independent check to our analytical results, we also solved the model
numerically for some large values of $N$. We used this solution to compute $1/N$ correction
for these large $N$ values, then extrapolate them to the $N\rightarrow\infty$ limit using Vanden Broeck 
and Schwartz algorithm. Our numerical results strongly support the analytical derivations presented here.
Spectral equivalences between diagonal-nondiagonal and diagonal-diagonal, 
nondiagonal-nondiagonal and diagonal-diagonal \cite{DNPR,DNPR2,Bajnok2} open XXZ spin chains
have been shown to exist. Hence, one may attempt to explain the diagonal (Dirichlet)
behaviour of the model studied here by some such equivalence. However, to our knowledge,
such equivalences have been found when the boundary parameters obey certain constraint \cite{Ne2}--\cite{YNZ},
which is not the case for the model we considered here, as already remarked in Footnote 1. 
Hence, the question about the ``Dirichlet-like'' behaviour remains for now. We hope to be able to 
resolve this issue soon.

There are many other open questions that one can explore and address further. 
For example, similar analysis involving boundary excitations can also be carried 
out. This can be really challenging even for the diagonal (Dirichlet) case \cite{AhnBellacosa,SkorikSaleur}. 
Further, solution for more general XXZ model involving multiple $Q(u)$ functions \cite{MNS,MNS3}, can 
also be utilized in similar capacity to explore these effects. Last but not least, excitations
due to other objects that we choose to ignore here, such as special roots/holes and 
so forth can also be explored for these models in order to make the study more
complete. We look forward to address some of these issues in near future.    

\section*{Acknowledgments}

I would like to thank R.I. Nepomechie for his invaluable advice, suggestions 
and comments during the course of completing this work. I also fully appreciate the
financial support received from the Department of Physics, University of Miami.

\end{document}